\documentclass{aa}
\usepackage{amssymb}
\usepackage{graphicx}
\usepackage{natbib}
\bibpunct{(}{)}{;}{a}{}{,}

\begin{document}
   \title{Drift instabilities in the solar corona within the multi-fluid description}


   \author{R. Mecheri
          \inst{1}
          \and
           E. Marsch\inst{1}  
           }

   \offprints{R. Mecheri}

   \institute{Max-Planck-Institut f\"{u}r Sonnensystemforschung,
              Max-Planck-Strasse 2, 37191 Katlenburg-Lindau, Germany\\
              \email{mecheri@mps.mpg.de}
             }

    \date{Received xxxxx; accepted xxxxx}


\abstract
{Recent observations revealed that the solar atmosphere is highly
structured in density, temperature and magnetic field. The presence
of these gradients may lead to the appearance of currents in the
plasma, which in the weakly collisional corona can constitute
sources of free energy for driving micro-instabilities. Such
instabilities are very important since they represent a possible
source of ion-cyclotron waves which have been conjectured to play a
prominent role in coronal heating, but whose solar origin remains
unclear.}
{Considering a density stratification transverse to the magnetic
field, this paper aims at studying the possible occurrence of
gradient-induced plasma micro-instabilities under typical
conditions of coronal holes.}
{Taking into account the WKB (Wentzel-Kramers-Brillouin)
approximation, we perform a Fourier plane waves analysis using the
collisionless multi-fluid model. By neglecting the electron inertia,
this model allows us to take into account ion-cyclotron wave effects
that are absent from the one-fluid model of magnetohydrodynamics
(MHD). Realistic models of density and temperature, as well as a 2-D
analytical magnetic-field model, are used to define the background
plasma in the open-field funnel in a polar coronal hole. The
ray-tracing theory is used to compute the ray path of the unstable
waves, as well as the evolution of their growth rates during the
propagation.}
{We demonstrate that in typical coronal hole conditions, and when
assuming typical transverse density length scales taken from radio
observations, the current generated by a relative electron-ion drift
provides enough free energy for driving the mode unstable. This
instability results from a coupling between oppositely propagating
slow-mode waves. However, the ray-tracing computation shows that the
unstable waves propagate upward to only a short distance but then
are reflected backward. The corresponding growth rate increases and
decreases intermittently in the upward propagating phase, and the
instability ceases while the wave is propagating downward.}
{Drift currents caused by fine density structures in the magnetically
open coronal funnels can provide sufficient energy for driving
plasma micro-instabilities, which constitute a possible source of
the ion-cyclotron waves that have been invoked for coronal heating.}

   \maketitle
%

\section{Introduction}\label{introduction}


New observations of coronal structures made by the high-resolution
ultraviolet, extreme ultraviolet and X-ray telescopes onboard the
SOHO and TRACE satellites have revealed the structure of the solar
corona as highly filamentary and inhomogeneous. In particular, TRACE
allowed observations of coronal structures be made down to spatial
scales smaller than 1000~km and provided evidence for a fine
structuring of coronal loops consisting of many individual threads
\citep{mecheri:Aschwanden00,mecheri:McEwan}. These loop filaments
are very thin with a thickness of the order of the spatial
resolution element of TRACE. Moreover, radio propagation
measurements have shown that the outer corona is also highly
inhomogeneous in the direction perpendicular to the magnetic field.
This nonuniformity appears in the form of filamentary ray-like
structures, extending radially from the coronal base into the
corona. The perpendicular length scale of the associated density
filaments can be as small as 1~km at the coronal base, and is about
10~km at 2--5~R$_{\odot}$ \citep{mecheri:Woo96,mecheri:Woo06}, which
is by about two orders of magnitude smaller than the observational
limit of TRACE. Strong inhomogeneity generally prevails in the lower
corona, and precisely at the boundaries between dilute open funnels
and dense closed loops, which can be associated with sharp gradients
in the background plasma quantities. Such gradients play a crucial
role in the theory of waves in nonuniform plasma. The interaction of
waves with plasma inhomogeneities brings important new physical
effects, such as dispersion, phase mixing \citep{mecheri:Heyvaerts1}
and resonant absorption \citep{mecheri:Ionson2}.

Additionally, if the plasma has gradients of the density,
temperature, pressure or magnetic field a plasma current may exist
and thus provide free energy for driving micro-instabilities. They
can be very important in the coronal context, since they may
constitute an affluent source of the high-frequency ion-cyclotron
waves that have been invoked to play a prominent role in coronal
heating through kinetic dissipation
\citep{mecheri:Kohl97,mecheri:Cranmer99a,mecheri:Marsch01}, but for
which the coronal origin still remains unclear.
\citet{mecheri:Axford92,mecheri:Axford95} suggested that
ion-cyclotron waves could originate in the lower corona from
small-scale reconnection events in the magnetic network. In the
extended corona ion-cyclotron waves may be locally generated through
a turbulent cascade of low-frequency MHD-type waves towards
high-frequency waves
\citep{mecheri:Hollweg86,mecheri:Tu88,mecheri:Isenberg90,mecheri:Marsch90a,mecheri:Marsch90b,mecheri:Hu99,mecheri:Li99,mecheri:Ofman02,mecheri:Markovskii06},
or via mode conversion driven by reflection or refraction of
low-frequency MHD waves \citep{mecheri:Matthaeus,mecheri:Cranmer05}.
Another possible scenario of local generation of ion-cyclotron waves
is by plasma micro-instabilities that are driven by an intermittent
electron heat flux and plasma outflows accompanying micro-flares
events
\citep{mecheri:Markovskii04a,mecheri:Markovskii04b,mecheri:Voitenko02},
by cross-field current fluctuations of low-frequency MHD modes
\citep{mecheri:Markovskii01,mecheri:Markovskii02,mecheri:Zhang03},
or by coronal ion beams \citep{mecheri:Mecheri07}.

The aim of this paper is to study the possible occurrence of
micro-instabilities in the ion-cyclotron frequency range, which are
induced by cross-field currents and occur under typical coronal-hole
conditions. In our present case, these currents are generated by an
ion-electron cross-field drift that is supported by a density
stratification perpendicular to the ambient magnetic field. Linear
mode analysis is used in the framework of a collisionless
multi-fluid model. By neglecting the electron inertia, this model
permits the consideration of ion-cyclotron wave effects that are
absent from the one-fluid MHD model. The ion-cyclotron wavelengths
are of the order of the ion inertial length,
$\lambda_{i}=V_{Ai}/\Omega_{i}=c/\omega_{i}$, i.e.,
$\lambda\approx\lambda_{i}$, where $V_{Ai}=B_0/\sqrt{\mu_{0} n_i
m_i}$ is the Alfv\'{e}n speed of the ion species $i$ with mass $m_i$
and density $n_i$, and $\Omega_{i}$ its cyclotron frequency,
respectively, $\omega_{i}$ plasma frequency, and where
$B_0(=\sqrt{B_{0x}^{2}+B_{0z}^{2}})$ denotes the background magnetic
field magnitude. In the weakly collisional corona, the length
$\lambda_{i}$ is much smaller than the electron-ion collisional mean
free path $\lambda_{ei}$, i.e., $\lambda_{i}\ll\lambda_{ei}$, and
consequently $\lambda\ll\lambda_{ei}$, which justifies the
collisionless limit adopted for this study.

Realistic models of the density and temperature, as well as a 2-D
funnel model describing an open magnetic field, are used to define
the background plasma. Considering the WKB approximation (i.e., the
wavelengths of interest are smaller than the non-uniformity length
scale), we first solve locally the dispersion relation and then
perform a non-local wave analysis using the ray-tracing theory. This
theory allows us to compute the ray path of the unstable waves in
the funnel as well as the spatial variation of their growth rate.

This paper is structured as follows. In Sect.\ref{background}, we
present the 2-D analytical funnel model used in this study to
represent an open-field region in a coronal hole. Then in
Sect.\ref{linear}, we describe how the local and non-local
(ray-tracing) linear perturbation analysis is carried out, using
the multi-fluid model. The results are presented and discussed in
Sect.\ref{results}, and finally we give our conclusions in
Sect.\ref{conclusion}.

\section{Background plasma configuration}\label{background}

\begin{figure}
\centering
\includegraphics[width=6.8cm]{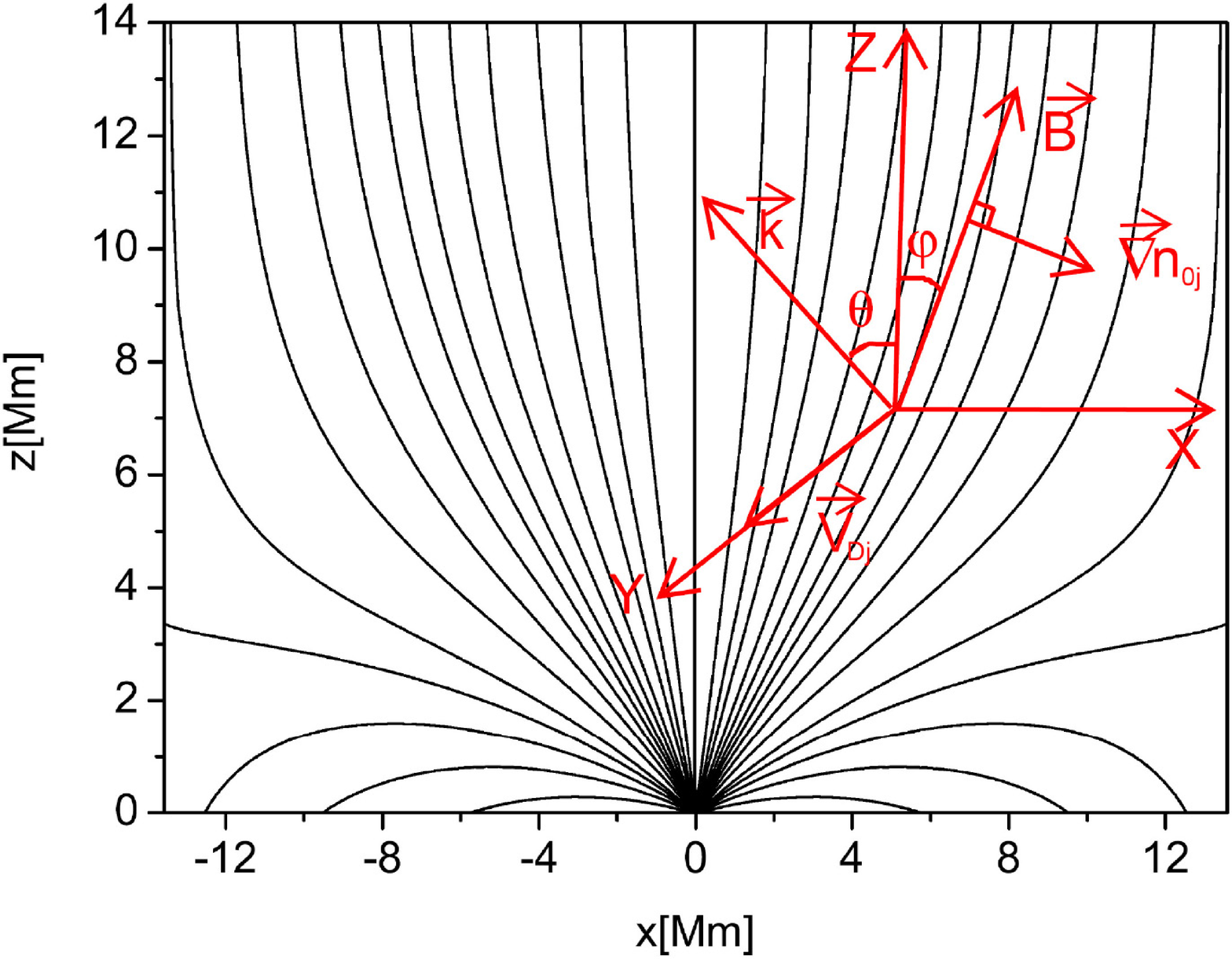}
\caption{Funnel magnetic field geometry as obtained from the 2-D
potential field model derived by \citet{Mecheri:Hackenberg}. The
field lines emerge from the boundary between two adjacent
supergranules ($x=0$) and expand rapidly to fill the corona. The
photospheric level is at $z=0$. Coordinate axes, wave vector and
drift velocity are shown in red.} \label{Figfunnel}
\end{figure}

For the background plasma density and temperature we use the model
of \citet{Mecheri:Fontenla} for the chromosphere and the model of
\citet{Mecheri:Gabriel} for the lower corona.
The 2-D potential-field and current-free funnel model derived by
\citet{Mecheri:Hackenberg} is used to define the background magnetic
field (Fig.\ref{Figfunnel}). Analytically, this model is given by:
{\setlength\arraycolsep{0.005em}
\begin{eqnarray}
B_{0x}(x,z)=\frac{(B_{max}-B_{00})D}{2\pi(D-d)}ln\frac{cosh\frac{2\pi
z}{D}- cos(\frac{\pi d}{D}+\frac{2\pi x}{D})}{cosh\frac{2\pi z}{D}-
cos(\frac{\pi d}{D}-\frac{2\pi x}{D})}\nonumber\\
\end{eqnarray}
\vspace{-0.7cm}
\begin{eqnarray}
B_{0z}(x,z)=B_{00}+(B_{max}-B_{00})\left[-\frac{d}{D-d}
+\frac{D}{(D-d)\pi}\right.\times\nonumber\\
\left(arctan\frac{cosh\frac{2\pi z}{D}~sin\frac{\pi d}{2D}+
sin(\frac{\pi d}{2D}+\frac{2\pi x}{D})}{sinh\frac{2\pi
z}{D}~cos\frac{\pi d}{2D}}+\right.
\nonumber\\
\left.\left.arctan\frac{cosh\frac{2\pi z}{D}~sin\frac{\pi d}{2D}+
sin(\frac{\pi d}{2D}-\frac{2\pi x}{D})}{sinh\frac{2\pi
z}{D}~cos\frac{\pi
d}{2D}}\right)\right]\nonumber\\
\end{eqnarray}}\\
with the relevant parameters given as follow: \textit{D}=30
\textrm{Mm}, \textit{d}=0.34 \textrm{Mm}, $B_{00}$=11.8 \textrm{G},
$B_{max}$=1.5 \textrm{kG}.

To model nonuniformity, we consider the length scales of the
background plasma density and pressure in the z-direction, \textit{H},
to be much bigger than the one in the x-direction, \textit{L}, i.e.,
$H=((1/n_{0j})\partial n_{0j}/{\partial z})^{-1}\gg
L=((1/n_{0j})\partial n_{0j}/{\partial x})^{-1}$. Thus only a
density gradient in the x-direction will be considered, for which
use the following analytical density model:
\begin{equation}
n_{0j}(x)=n_{0j}(1+\frac{x}{L}),
\label{nx}
\end{equation}
whereby it is assumed that the density increases linearly in the
x-direction, but only locally within the studied region where it is
getting bigger by a factor 10. This assumption avoids a large growth
of the density while the x-coordinate increases.

\section{Linear perturbation analysis}\label{linear}
The fluid equations associated with the particle species $j$ are
given by:
\begin{eqnarray}
  \frac{\partial n_{j}}{\partial
  t}+\nabla\cdot(n_{j}\textbf{v}_{j})=0,
  \label{mass}
\end{eqnarray}
\begin{equation}
  m_{j}n_{j}(\frac{\partial \textbf{v}_{j}}{\partial
  t}+\textbf{v}_{j}\cdot\nabla \textbf{v}_{j})+
  \nabla p_{j}-q_{j}n_{j}(\textbf{E}+\textbf{v}_{j}\times
  \textbf{B})=0,
  \label{momentum}
\end{equation}
\begin{equation}
  \frac{\partial p_{j}}{\partial
  t}+\textbf{v}_{j}\cdot\nabla p_{j}+\gamma_{j}
  p_{j}\nabla\cdot\textbf{v}_{j}=0,
  \label{energy}
\end{equation}
where $m_{j}$, $n_{j}$, $\textbf{v}_{j}$, $p_{j}$ and $\gamma_{j}$
are respectively the mass, density, velocity, pressure (which for
simplicity is here assumed to be isotropic) and the polytropic index.
The subscript $j$ stands for electron $e$, proton $p$
or alpha particle $\alpha$ (He$^{2+}$). The electric field
$\textbf{E}$ and the magnetic field $\textbf{B}$ are linked through
Faraday's law:
\begin{equation}
\nabla\times \textbf{E}=-\frac{\partial \textbf{B}}{\partial t}.
\end{equation}

\subsection{Linearization procedure}

We express all the quantities in the above equations as a sum of an
unperturbed stationary part (with subscript 0) and a perturbed part
(with subscript 1) much smaller than the stationary part as follows:
\begin{eqnarray}
\hspace{-0.3cm}&&n_{j}=n_{0j}(x)+n_{1j},~p_{j}=p_{0j}(x)+p_{1j},
\nonumber\\[0.2cm]
\hspace{-0.3cm}&&\textbf{v}_{j}=\textbf{v}_{Dj}+\textbf{v}_{1j},
~\textbf{B}=\textbf{B}_{0}(x,z)+\textbf{B}_{1},
~\textbf{E}=\textbf{E}_{0}+\textbf{E}_{1},\nonumber\\[0.2cm]
\hspace{-0.3cm}&& n_{1j}\ll n_{0j}, p_{1j}\ll p_{0j},
\left|\textbf{v}_{1j}\right| \ll \left|\textbf{v}_{Dj}\right|,
\left|\textbf{B}_{1}\right|\ll \left|\textbf{B}_{0}\right|,
\label{perturbation}
\end{eqnarray}
where we have considered charge neutrality in the unperturbed
stationary plasma, i.e., $\sum_{j}q_{j}n_{0j}=0$. Due to density
stratification in the x-direction, a current is carried by drifting
electrons and ions in opposite directions parallel to the y-axis,
with a drift velocity given by:

\begin{equation}
\textbf{v}_{Dj}= - \frac{1}{L}\frac{C_{sj}^{2}}{\gamma_{j}\Omega_{j}}
\frac{B_{0}}{B_{0z}}\textbf{\^{y}} \label{VDJ},
\end{equation}
where $C_{sj}=\sqrt{\gamma_{j}k_{B}T_{j}/m_{j}}$ and
$\Omega_{j}=q_{j}B_{0}/m_{j}$ are, respectively, the acoustic speed
and the cyclotron frequency of the species $j$, and $k_{B}$ is the
Boltzman constant.

The zero-order terms cancel out when Eq.(\ref{perturbation}) is
inserted into the multi-fluid Eqs.(\ref{mass})-(\ref{energy}).
Neglecting the products of first-order nonlinear terms, we get a
system of linear equations:
\begin{equation}
i(\omega-\textbf{k}\cdot\textbf{v}_{Dj})\frac{n_{1j}}{n_{0j}}
-\textbf{v}_{1j}\cdot\frac{\nabla n_{0j}}{n_{0j}}
-i\textbf{k}\cdot\textbf{v}_{1j}=0,
\end{equation}
\begin{eqnarray}
i(\omega-\textbf{k}\cdot\textbf{v}_{Dj})\textbf{v}_{1j}&+&
\frac{\Omega_{j}}{\left|\textbf{B}_{0}\right|}
\left(\textbf{E}_{1}+\textbf{v}_{1j}\times\textbf{B}_{0}
+\textbf{v}_{Dj}\times\textbf{B}_{1}
\right)\nonumber\\
&-&C_{sj}^{2}\left(i
\frac{p_{1j}}{p_{0j}}\textbf{k}-\frac{n_{1j}}{n_{0j}} \frac{\nabla
p_{0j}}{p_{0j}}\right)=0,
\end{eqnarray}
\begin{equation}
i(\omega-\textbf{k}\cdot\textbf{v}_{Dj})\frac{p_{1j}}{p_{0j}}-
\textbf{v}_{1j}\cdot\frac{\nabla p_{0j}}{p_{0j}}
-i\gamma_{j}\textbf{k}\cdot\textbf{v}_{1j}=0,
\end{equation}
where all the perturbed quantities have been expressed in terms of
the amplitudes of plane waves. Fourier analysis has turned all
derivatives into factors: $\partial/\partial t\rightarrow -i \omega$
and $\nabla\rightarrow i \textbf{k}$, where $\omega$ is the wave
frequency and $\textbf{k}$ the wave vector. Note that in Eq. (11)
the term $\textbf{v}_{1j}\cdot\nabla \textbf{v}_{Dj}$ has been
neglected in comparison with the term
$(\textbf{k}\cdot\textbf{v}_{Dj})\textbf{v}_{1j}$ since it is
smaller by the ratio $2\pi L_{B_{0z}}/\lambda$ where
$L_{B_{0z}}=(|\nabla{B}_{0z}|/B_{0z})^{-1}$ is the length scale of
the z-component of the background magnetic field and $\lambda$ the
wave length of interest, i.e.:
\begin{eqnarray}
\textbf{v}_{1j}\cdot\nabla
\textbf{v}_{Dj}/(\textbf{k}\cdot\textbf{v}_{Dj})\textbf{v}_{1j}
&&\approx L_{B_{0z}}^{-1}/(2\pi/\lambda)\nonumber\\
&&\approx \lambda/(6L_{B_{0z}})\approx10^{-5}\ll1.
\end{eqnarray}
%
%
In this context we want to mention, and remind the reader of, the
classical guiding-center drift caused by the gravity force
($\textbf{g}$ is the gravitational acceleration) in a magnetic
field, i.e.:
\begin{equation}
\textbf{v}_{gj} = \frac{1}{\Omega_j} \textbf{g} \times
\frac{\textbf{B}}{ \mid \textbf{B}\mid}.
\end{equation}
In the corona this drift would lead to a current perpendicular to
the magnetic field, and could thus also provide free energy which,
as was shown in considerable detail by \citet{mecheri:Brinca02,
mecheri:Brinca03}, could lead to various kinds of plasma
instabilities and wave excitation. In comparison with
Eq.(\ref{VDJ}), we obtain as an order-of-magnitude estimate for the
ratio of these two drift speeds the result:
\begin{equation}
\frac{ v_{gj}}{ v_{Dj}} \approx \frac{g}{\Omega_j}\frac{\Omega_j
L}{C_{sj}^{2}} \approx \frac{L}{H},
\end{equation}
where we used again the gravitational scale height $H$ as defined in
the previous section for the background density being thermally
stratified in the z-direction. So the small horizontal density
variation considered here yields a larger drift than the one induced
by vertical gravity. However, transverse currents driven by
gravitational stratification may play a role in generating plasma
instability at the gyrokinetic scale, but the growth rate is
expected to be smaller by the ratio $H/L$.

\subsection{Dispersion relation}

The above linearized equations are combined in order to obtain a
linear relation between the current density $\textbf{J}_{1}$ and
the electric field $\textbf{E}_{1}$:

\begin{equation}
\textbf{J}_{1}=\vec{\sigma}\cdot\textbf{E}_{1},
\end{equation}
where $\vec\sigma$ is the conductivity tensor which is related to
the dielectric tensor $\vec\epsilon$ through the following relation:

\begin{equation}
\vec\epsilon=\textbf{I}+\frac{i}{\omega\varepsilon_{0}}\vec\sigma.
\label{dielectric}
\end{equation}
%
The dispersion relation is obtained using electrodynamics theory
\citep[e.g.,][]{Mecheri:Stix}:
\begin{equation}
D(\omega,\textbf{k},\textbf{r})=\textrm{Det}\left[\frac{c^{2}}
{\omega^{2}}
\textbf{k}\times(\textbf{k}\times\textbf{E})+\vec{\epsilon}(\omega,
\textbf{k},\textbf{r})\cdot\textbf{E}\right]=0,
\label{dispersion}
\end{equation}
where $c$ is the speed of light in vacuum and $\textbf{r}$ is the
large-scale position vector.
%

\subsection{Ray-tracing equations}\label{RT}

Considering the WKB approximation, the ray-tracing problem consists
in solving a system of ordinary differential equations of the
Hamiltonian form \citep{Mecheri:Weinberg,Mecheri:Lighthill}. These
equations, that solve for the wave  frequency $\omega$, the wave
vector $\textbf{k}$ and the space coordinate $\textbf{r}$, have been
formulated by \citet{Mecheri:Bernstein}, and in the simple case of a
Hermitian dielectric tensor are given by:
\begin{equation}
\frac{\textrm{d}\omega}{\textrm{d}t}=-\frac{\partial
D(\omega,\textbf{k} ,\textbf{r})/\partial t}{\partial
D(\omega,\textbf{k} ,\textbf{r})/\partial \omega}=0,\label{rt1}
\end{equation}
\begin{equation}
\frac{\textrm{d}\textbf{k}}{\textrm{d}t}=~\frac{\partial
D(\omega,\textbf{k} ,\textbf{r})/\partial \textbf{r}}{\partial
D(\omega,\textbf{k} ,\textbf{r})/\partial \omega},\label{rt2}
\end{equation}
\begin{equation}
\frac{\textrm{d}\textbf{r}}{\textrm{d}t}=-\frac{\partial
D(\omega,\textbf{k} ,\textbf{r})/\partial \textbf{k}}{\partial
D(\omega,\textbf{k} ,\textbf{r})/\partial \omega}\cdot\label{rt3}
\end{equation}
A generalization of these equations to the case of an anti-Hermitian
dielectric tensor was also proposed by \citet{Mecheri:Bernstein}. In
this case, additionally to the ray path, the growth rate of the
instability can also be computed. Note that Eq.(\ref{rt1}) is set to
zero since the dispersion relation does not depend explicitly on the
time $t$ (the background plasma is stationary). The above
differential equations represent a set of initial values problem
which can be solved using initial conditions obtained from the local
solutions of the dispersion relation (Eq.\ref{dispersion}).

\section{Numerical results}\label{results}
\subsection{Local stability analysis}

The dispersion relation Eq.(\ref{dispersion}) is solved numerically
for the funnel location (x=9.4 Mm, z=2.5 Mm) characterized by a
magnetic field inclination angle of $\varphi=85.3^{\circ}$ with
respect to the normal on the solar surface. The value of the density
length scale in the x-direction, \textit{L}, is chosen to be 1~km
according to results obtained from radio propagation measurements
\citep{mecheri:Woo96,mecheri:Woo06}. We consider the cases of a
two-fluid (e-p) and a three-fluid (e-p-He$^{2+}$) model, where we
consider the effect a second ion population of alpha particles
(He$^{2+}$) with typical coronal abundance. First, the local
solutions of the dispersion relation Eq.(\ref{dispersion}) are
presented, and then the results obtained from the non-local
ray-tracing equations.

\subsubsection{Two-fluid (e-p) drift-plasma}
\begin{figure}
\begin{center}
\large\mbox{x=9.4~Mm,~z=2.5~Mm,~
$\varphi\approx85.3^{\circ}$,~$\theta=85^{\circ}$}
\includegraphics[width=7.6cm]{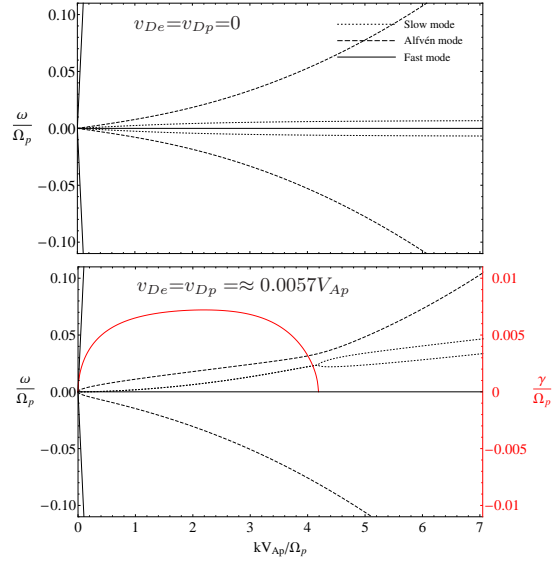}
\end{center}
\vspace{-0.5cm}\caption{Two-fluid dispersion curves at the funnel
location (x=9.4 Mm, z=2.5 Mm) with a \textbf{B}$_{0}$-inclination
angle $\varphi=85.3^{\circ}$. The angle of propagation is
$\theta=85^{\circ}$. \textbf{Top}: The case of a uniform plasma
density, i.e. $v_{Dj}=0$, with a plasma beta
$\beta_{e}=\beta_{p}\approx0.0097$. \textbf{Bottom}: In the presence
of a drift due to a density gradient with a scale length of
\textit{L}=1~km. Here $\omega$ and $k$ are normalized, respectively,
to the proton cyclotron frequency, $\Omega_{p}$, and the proton
inertial length, $\Omega_{p}/V_{Ap}$, where
$V_{Ap}=B_{0}/\sqrt{\mu_{0}n_{0p}m_{p}}$ is the proton Alfv\'{e}n
speed ($\mu_{0}$ is the magnetic permeability in vacuum). Here
$T_{0e}=T_{0p}$.} \label{Dr2-2D-d}
\end{figure}

The two-fluid drift-plasma configuration consists of a plasma made
of electrons with density $n_{0e}$ and protons ($n_{0p}$) in
relative drift to each other in opposite directions along the y-axis
and perpendicular to the background magnetic field with velocities
given by Eq.(\ref{VDJ}). The drift velocity is sustained by the
presence of a local density gradient in the x-direction, as
described by Eq.(\ref{nx}). For the purpose of comparison, the
dispersion curves in the case of a vanishing density gradient,
\textit{L}$^{-1}$=0, and thus no relative drift between electrons
and ions, i.e., $v_{De}$=$v_{Dp}$=0, are given in the top panel of
Fig.\ref{Dr2-2D-d}, corresponding to an angle of propagation
$\theta=85^{\circ}$. Consequently, there is no free energy in the
plasma, and all the modes remain stable. In this case the dispersion
relation Eq.(\ref{dispersion}) is quadratic and three stable modes
are present. Each one is represented by an oppositely propagating
($\omega>0$ and $\omega<0$) pair of waves. These modes represent the
extensions of the usual Slow (dotted line), Alfv\'{e}n (dashed
line), and Fast (solid line) MHD modes into the high-frequency
domain around $\omega=\Omega_{p}$ (=$eB_{0}/m_{p}$), where these
waves become dispersive.

However, in the case \textit{L}=1 km, the relative drift between
electrons and protons is not zero, and the dispersion curves are
strongly modified with a breaking of the symmetry between forward
($\omega>0$) and backward ($\omega<0$) propagating waves, as it is
shown in the bottom panel of Fig.\ref{Dr2-2D-d}. This behavior
affects particularly the slow mode. Indeed, the free energy provided
by the relative drift between electrons and protons, which is
sustained by the density gradient, leads to the appearance of
regions of instability resulting from the coupling between forward
and backward propagating slow modes. These two single initially
stable modes couple into one single unstable mode, for which the
normalized growth rate $\gamma/\Omega_{p}$ is given by the red
curve, which does not exceed 0.007.

\begin{figure*}
\begin{center}
\Large\mbox{x=9.4~Mm,~z=2.5~Mm, ~$\varphi\approx85.3^{\circ}$}\\
\includegraphics[width=12cm]{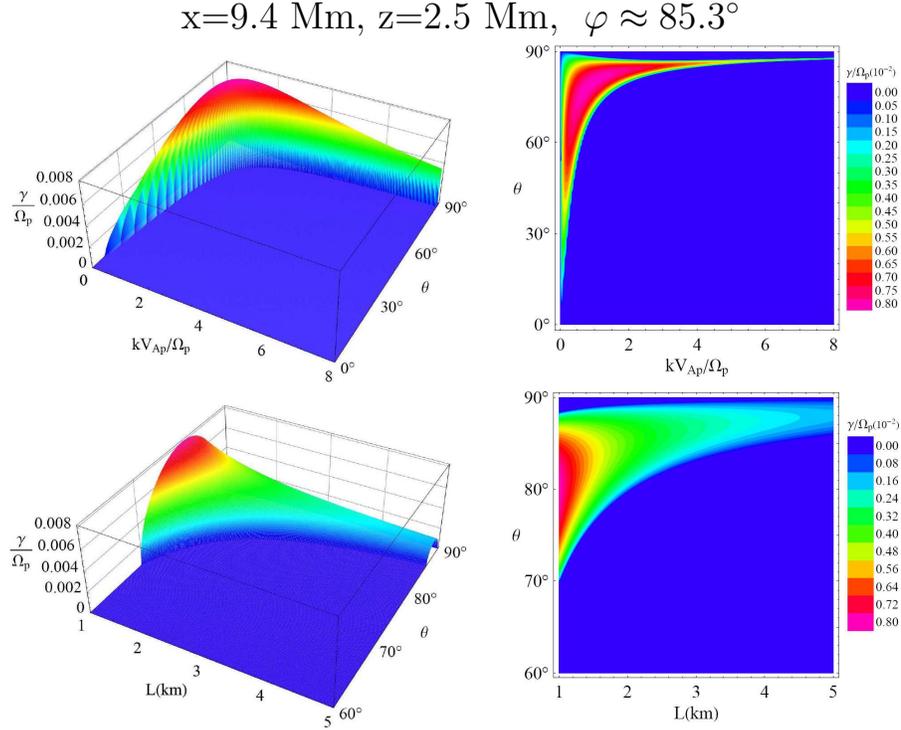}
\end{center}
\vspace{-0.5cm}\caption{Growth rate of the two-fluid instability
resulting form the coupling between forward and backward propagating
slow modes at the funnel location (x=9.4 Mm, z=2.5 Mm) with a
\textbf{B}$_{0}$-inclination angle of $\varphi=85.3^{\circ}$.
\textbf{Top}: As a function of the angle of propagation $\theta$ and
the normalized wave number, $kV_{Ap}$/$\Omega_{p}$, for a density
length-scale in the x-direction of \textit{L}=1~km
($v_{Dp}=v_{De}\approx0.0057V_{Ap}$). \textbf{Bottom}: As a function
of $\theta$ and \textit{L} for $kV_{Ap}/\Omega_{p}=1$.}
\label{Dr2-3D}
\end{figure*}

As it is shown in the top panel of Fig.\ref{Dr2-3D}, for small angle
of propagation, this instability is very weak and concentrated at
small wave number $kV_{Ap}/\Omega_{p}\ll 1$, and extends to larger
\textit{k} with sensitively bigger growth rate $\gamma$ at large
angle of propagation $\theta$. This instability is also shown in the
bottom panel of Fig.\ref{Dr2-3D} as a function of the propagation
angle $\theta$ and the length scale of the density gradient
\textit{L} in the x-direction at the location (x=9.4 Mm, z=2.5 Mm).
It is clearly seen, that with increasing \textit{L}, the instability
growth rate decreases and covers gradually a smaller range in the
angle of propagation $\theta$. The maximum growth rate is
$\gamma/\Omega_{p}\approx0.008$ for \textit{L}=1~km and for $\theta$
between 75$^{\circ}$ and 85$^{\circ}$. For larger \textit{L}, the
growth rate decreases and the instability appears only at large
angles of propagation ($\theta\gtrsim85^{\circ}$). This result is
expected since the larger the length scale of density \textit{L} is
the more decreases the particles drift speed, by virtue of
(Eq.\ref{VDJ}), providing thus less free energy to the plasma for
driving the instability.

\subsubsection{Three-fluid (e-p-He$^{2+}$) drift-plasma}

\begin{figure}
\begin{center}
\large\mbox{x=9.4~Mm,~z=2.5~Mm,~
$\varphi\approx85.3^{\circ}$,~$\theta=85^{\circ}$}
\includegraphics[width=7.6cm]{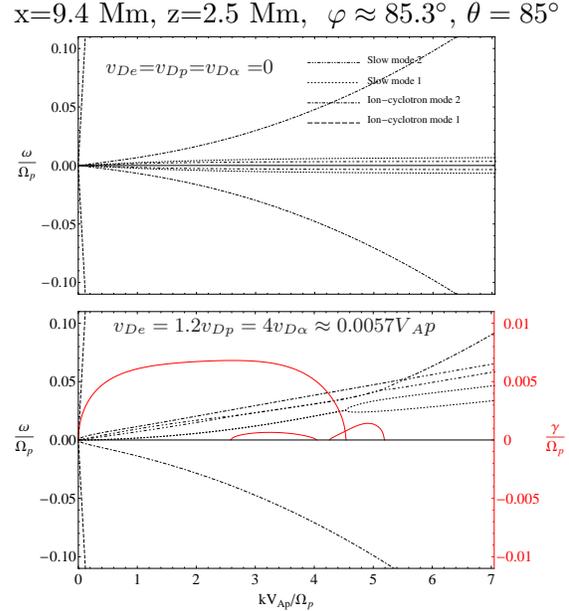}
\end{center}
\vspace{-0.5cm}\caption{Three-fluid dispersion curves at the
location (x=9.4 Mm, z=2.5 Mm) with a \textbf{B}$_{0}$-inclination
angle $\varphi=85.3^{\circ}$. The angle of propagation is
$\theta=85^{\circ}$. \textbf{Top}: The case of a uniform plasma,
i.e., $v_{Dj}=0$, with the plasma-beta
$\beta_{e}=1.2\beta_{p}=4\beta_{\alpha}\approx0.0097$.
\textbf{Bottom}: In the presence of a drift due to a density
gradient with a length scale \textit{L}=1~km.
Here $T_{0\alpha}=3T_{0p}=3T_{0e}$.}
\label{Dr3-2D-d}
\end{figure}

In the three-fluid drift-plasma configuration, in addition to
electrons ($n_{0e}$) and protons ($n_{0p}$), we consider the
presence of a second population of ions, namely alpha particles
He$^{2+}$ (indicated by $\alpha$) with a typical coronal abundance
of $n_{0\alpha}=0.1n_{0p}$. The electrons and ions are in opposite
relative drift according to Eq.(\ref{VDJ}) and caused by the
presence of a density gradient in the x-direction as in
Eq.(\ref{nx}).

For the purpose of comparison, the dispersion curves in the case of
\textit{L}$^{-1}$=0 (zero gradient of density) and consequently
$v_{De}$=$v_{Dp}$=$v_{D\alpha}$=0, are again given in the top panel
of Fig.\ref{Dr3-2D-d}, corresponding to the location (x=9.4 Mm,
z=2.5 Mm) with $\varphi=85.3^{\circ}$ and to an angle of propagation
$\theta=85^{\circ}$. In this case the dispersion relation
Eq.(\ref{dispersion}) is quadratic and the dispersion diagrams show
the presence of five stable modes. Each one of them is represented
by an oppositely propagating ($\omega>0$ and $\omega<0$) pair of
waves. We note the presence of: slow modes 1 and 2 (respectively
dotted and dashed-dot-dot lines), intermediate modes 1 and 2 (dashed
and dashed-dot lines), and one fast mode which is not shown here
because it presents a cut-off frequency around
$\omega_{co}\approx0.6\Omega_{p}$, which is out of the frequency
range presented in this study \citep[for further details
see][]{Mecheri:Mecheri}. Consequently, the consideration of a second
population of ions leads to the appearance of an additional slow and
ion-cyclotron mode.

\begin{figure*}
\begin{center}
\Large\mbox{x=9.4~Mm,~z=2.5~Mm, ~$\varphi\approx85.3^{\circ}$}\\
\includegraphics[width=12cm]{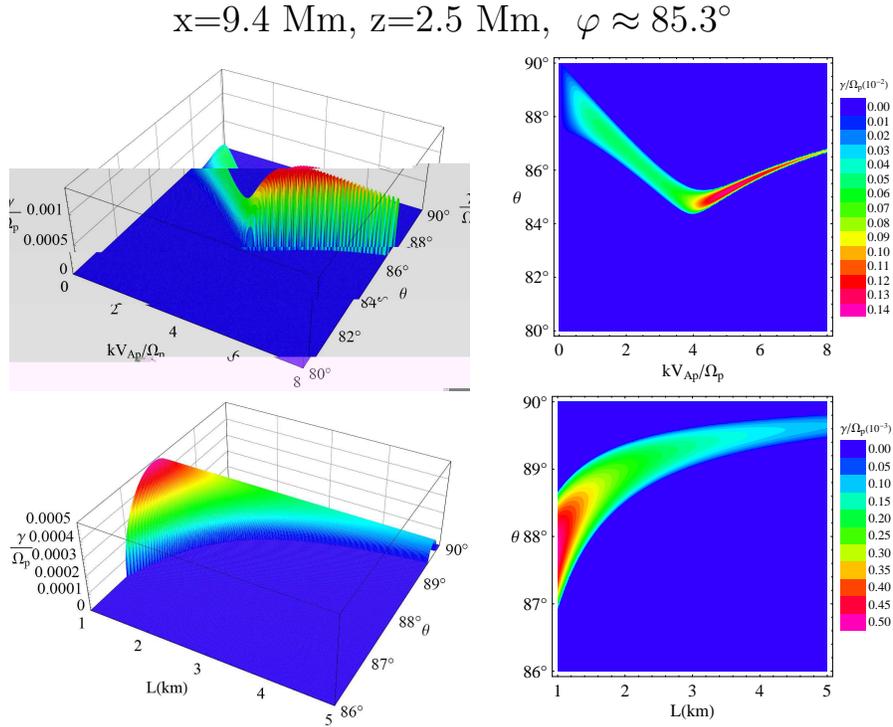}
\end{center}
\vspace{-0.5cm}\caption{Growth rate of the three-fluid slow mode 2
instability which is resulting from the coupling between forward and
backward propagating slow modes and is absent in the two-fluid case.
The results are given for the funnel location (x=9.4 Mm, z=2.5 Mm)
with a \textbf{B}$_{0}$-inclination angle of $\varphi=85.3^{\circ}$.
\textbf{Top}: As a function of the angle of propagation $\theta$ and
the normalized wave number $kV_{Ap}$/$\Omega_{p}$ for a
density-length scale in the x-direction of \textit{L}=1~km
($v_{De}=1.2v_{Dp}=4v_{D\alpha}\approx0.0057V_{Ap}$).
\textbf{Bottom}: As a function of $\theta$ and \textit{L} for
$kV_{Ap}/\Omega_{p}=1$.} \label{Dr3-3D-SS2}
\end{figure*}

In the case a non-zero density gradient in the $x$-direction
characterized by a length scale \textit{L}=1~km, the dispersion
relation is not quadratic anymore, and the symmetry between forward
and backward propagating modes is broken (bottom panel of
Fig.\ref{Dr3-2D-d}). Indeed, the free energy provided by the
relative drift between electrons and ions sustained by the density
gradient leads to the appearance of regions of instability. The
first instability is similar to the one found in the two-fluid model
It results from the coupling between forward and backward
propagating slow modes 1.
%
The second instability results from the coupling between forward and
backward propagating slow modes 2 with a sensitively smaller growth
rate, $\gamma\lesssim0.0015\Omega_{p}$ as compared to the first one.
%
%
This instability appears in general at very large angle of
propagation, i.e. $\theta\gtrsim84^{\circ}$ (top panel of
Fig.\ref{Dr3-3D-SS2}). As $\theta$ increases, the growth rate
appears in two different ranges of the wavenumber \textit{k}, the
first at small \textit{k}, i.e. $kV_{Ap}/\Omega_{p}\lesssim3$, and
the second at larger \textit{k}, i.e. $kV_{Ap}/\Omega_{p}\gtrsim5$.
The instability growth rate in general increases sensitively between
$kV_{Ap}/\Omega_{p}\approx4$ and 8, and has its maximum for
$kV_{Ap}/\Omega_{p}\approx5$. However, it is approximately 6 times
smaller than the slow-mode-1 maximum instability growth rate.
When plotted as a function of the propagation angle $\theta$ and the
length scale of the density gradient \textit{L} (bottom panel of
Fig.\ref{Dr3-3D-SS2}), this instability mainly appears at large
angle of propagation, i.e. $\theta \gtrsim87^{\circ}$, and the
corresponding growth rate decreases with increasing \textit{L}. The
maximum growth rate occurs for \textit{L}=1 km and for a propagation
angle around 88$^{\circ}$.


\subsection{Non-local stability analysis}

In this section we perform a non-local wave study using the
ray-tracing equations presented in Sect.(\ref{RT}). These equations
are computed using initial conditions obtained from the local
solutions of the dispersion relation Eq.(\ref{dispersion}) at the
starting location (x$_{0}$=9.4 Mm, y$_{0}$=0 Mm, z$_{0}$=2.5 Mm).
We consider both the two-fluid (e-p) and the three-fluid
(e-p-He$^{2+}$) case with a constant concentration of the alpha
particles (He$^{2+}$), i.e. $n_{\alpha}=0.1n_{p}$ along the ray
path. The density length scale in the x-direction \textit{L} is
chosen to be 1~km.

\subsubsection*{Two-fluid (e-p) drift-plasma}

\begin{figure*}
\begin{center}
\Large\mbox{x$_{0}$=9.4~Mm,~y$_{0}$=0,~z$_{0}$=2.5~Mm,
~$\varphi\approx85.3^{\circ}$}\\
\includegraphics[width=12cm]{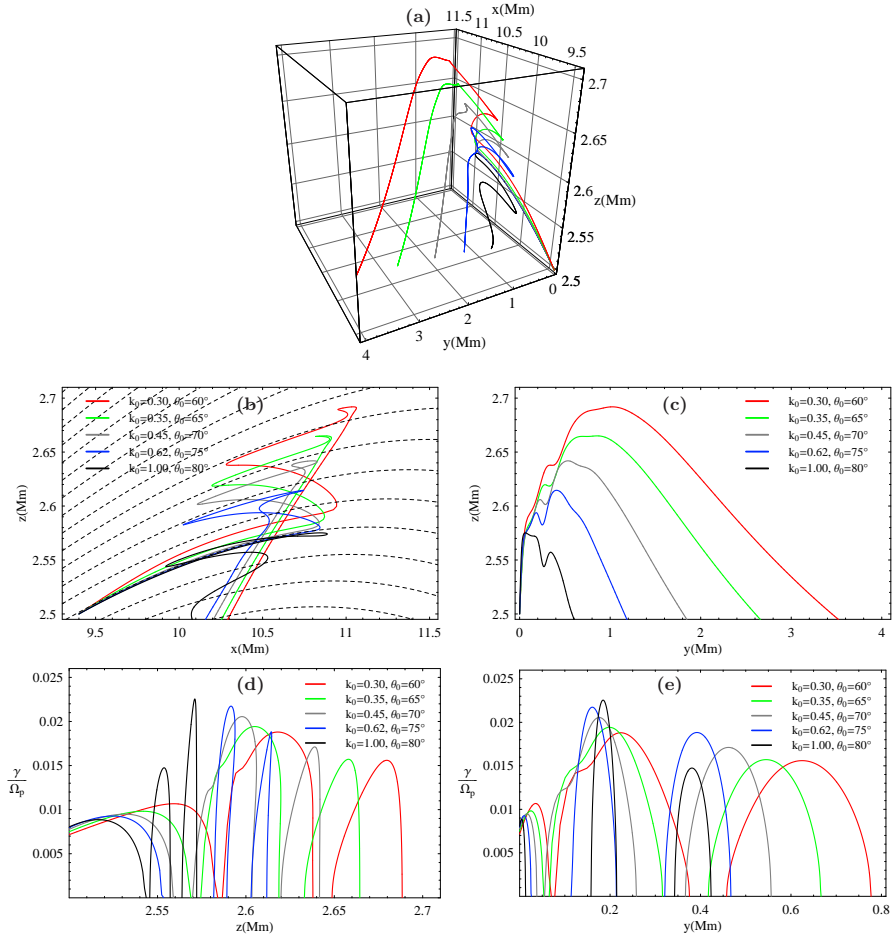}
\end{center}
\vspace{-0.5cm}\caption{Ray trajectory and growth rate of the
two-fluid slow modes instability. The waves are launched at the
initial location (x$_{0}$=9.4~Mm, y$_{0}$=0, z$_{0}$=2.5~Mm) in the
coronal funnel (with a \textbf{B}$_{0}$-inclination angle of
$\varphi_{0}\approx85.3^{\circ}$) and for different initial angle of
propagation $\theta_{0}$ and normalized wave number $k_{0}$, chosen
for each $\theta_{0}$ at the maximum growth rate. The dashed lines
represent the funnel magnetic field lines. (a) 3-D ray path; (b)
Projection of the ray path on the x-z plane; (c) Ray path projected
on the y-z plane; (d) Growth rate as a function of the z-coordinate;
(e) Growth rate as a function of the y-coordinate.}
\label{RTd23-SS}
\end{figure*}

As presented in the previous section, in the two-fluid case the
local solution of the dispersion relation showed the presence of an
instability resulting from the coupling between forward and backward
propagating slow modes. The ray path of this unstable wave, as well
as the spatial variation of its growth rate $\gamma$, when the wave
is launched at the initial location (x$_{0}$=9.4~Mm, y$_{0}$=0,
z$_{0}$=2.5~Mm), are presented on Fig.\ref{RTd23-SS}. The results
are shown for different initial angle of propagation $\theta_{0}$,
with which a different initial normalized wave number $k_{0}$ is
associated that is chosen to always correspond to the maximum growth
rate $\gamma_{max}$. The wave starts propagating from its initial
location and is well guided along the field lines up to
approximately the location x$\approx$11~Mm and z$\approx$2.6~Mm,
where its motion starts to become unguided and irregular (see
Fig.\ref{RTd23-SS}a, \ref{RTd23-SS}b and \ref{RTd23-SS}c). This
unguided motion continues until the wave reaches a maximum height
where it is reflected and starts to propagate downward almost in the
(y,z) plane to reach again its initial height, i.e., z=2.5~Mm. This
maximum height reached by the unstable wave is larger for smaller
initial angle of propagation $\theta_{0}$. Additionally, the smaller
$\theta_{0}$ is the more deeply can the wave propagate in the
y-direction. The corresponding instability growth rate $\gamma$ is
very irregular, and increases and decreases intermittently along the
ray path (Fig.\ref{RTd23-SS}d and \ref{RTd23-SS}e). The growth rate
$\gamma$ shows the appearance of peaks of maximum instability, which
are more important for large initial propagation angle $\theta_{0}$
but do not exceed $\gamma=0.025\Omega_{p}$. However, the instability
growth ceases at approximately the maximum height reached by the
wave, from where on it starts propagating downward.

\subsubsection*{Three-fluid (e-p-He$^{2+}$) drift-plasma configuration}

\begin{figure*}
\begin{center}
\Large\mbox{x$_{0}$=9.4~Mm,~y$_{0}$=0,~z$_{0}$=2.5~Mm,
~$\varphi\approx85.3^{\circ}$}\\
\includegraphics[width=12cm]{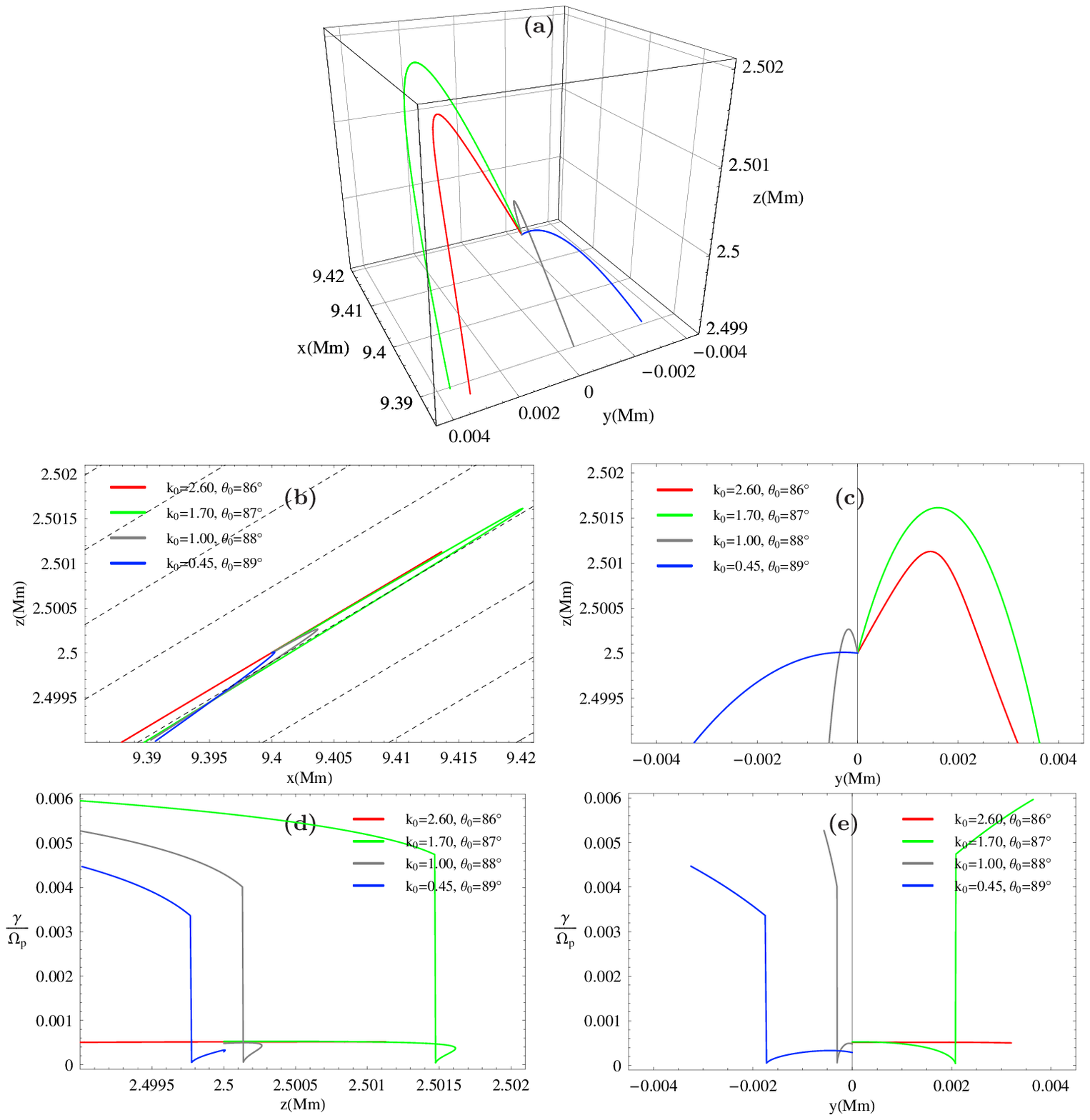}
\end{center}
\vspace{-0.5cm}\caption{Ray trajectory and growth rate of the
three-fluid slow mode 2 instability which is absent from the
two-fluid model. The waves are launched at the initial location
(x$_{0}$=9.4~Mm, y$_{0}$=0, z$_{0}$=2.5~Mm) in the coronal funnel
(with a \textbf{B}$_{0}$-inclination angle of
$\varphi_{0}\approx85.3^{\circ}$) and for different initial angle of
propagation $\theta_{0}$ and normalized wave number $k_{0}$, chosen
for each $\theta_{0}$ at the maximum growth rate. The dashed lines
represent the funnel magnetic field lines. (a) 3-D ray path; (b)
Projection of the ray path on the x-z plane; (c) Ray path projected
on the y-z plane; (d) Growth rate as a function of the z-coordinate;
(e) Growth rate as a function of the y-coordinate.}
\label{RTd23-SS2}
\end{figure*}

In this three-fluid model case, and at the location (x=9.4~Mm,
z=2.5~Mm), the solution of the dispersion relation shows the
presence of two instabilities both resulting from the coupling
between forward and backward propagating slow modes.
%
%
For the first instability, which is similar to the one obtained in
the case of the two-fluid model, the ray path as well as the spatial
variation of its growth rate $\gamma$ show a similar behavior as in
the two-fluid case.
%
%
The ray path as well as the growth rate variation of the second
instability, when the wave is launched at the initial location
(x$_{0}$=9.4~Mm, y$_{0}$=0, z$_{0}$=2.5~Mm), are presented in
Fig.\ref{RTd23-SS2}. The results show that the unstable wave starts
to propagate as a guided mode for a very small distance but then is
rapidly reflected to propagate downward (Fig.\ref{RTd23-SS2}a,
\ref{RTd23-SS2}b and \ref{RTd23-SS2}c). Additionally, the waves
propagates only to a very short distance in the y-direction, i.e.
$y<0.004$~Mm (Fig.\ref{RTd23-SS2}c).


\section{Conclusion}\label{conclusion}

We have studied gradient-drift electromagnetic instabilities in a
coronal funnel, using a collisionless two-fluid (e-p) and
three-fluid (e-p-He$^{2+}$) plasma model with finite pressure. While
neglecting electron inertia, this model allows us to take
ion-cyclotron wave effects into account. We have considered a
density stratification transverse to the ambient magnetic field with
the typical small length scales suggested by the observations.

First, a local perturbation analysis has been performed, whereby
the dispersion relation has been solved for a local region in a
coronal funnel. By comparison with the results obtained for a
uniform-density plasma, it was found that the dispersion curves are
strongly modified. For certain ranges of the wave number, two
initially stable modes merge into a single unstable mode. Indeed,
the free energy provided by the relative drift between the different
plasma species due to the density gradient leads to the appearance
of regions of instability. In the two-fluid model, this instability
results from the coupling between forward and backward propagating
slow modes. For large angles of propagation this instability extends
over a wide range in wave number and is characterized by a small
phase speed and an increasing growth rate as the angle of
propagation (with respect to the normal on the solar surface)
increases.

In addition, the consideration of a second ion population of alpha
particles (He$^{2+}$) with typical coronal abundance led to the
appearance of a new instability which also results from the coupling
of oppositely propagating slow mode waves. However, this second
instability in general covers only a small range in the wavenumber
domain and exists mainly for very large angles of propagation
$\gtrsim 82^{\circ}$, and it has a substantially smaller growth rate
by approximately a factor 6 as compared to the first instability.

The non-local analysis, which has been performed using ray-tracing
equations, revealed that the unstable waves, from their launching
point, start to propagate upward for a very small distance but then
are reflected and propagate downward. During this propagation, the
corresponding instability growth rate has a variable character and
is increasing, decreasing and vanishing intermittently.

Consequently, drift currents caused by the fine structuring of the
density in the magnetically open funnels of a coronal hole can
provide enough energy for driving plasma micro-instabilities. They
may in turn constitute a prolific source of the high-frequency
ion-cyclotron waves that have been invoked to play, through kinetic
wave dissipation, a prominent role in heating the open corona.

\bibliographystyle{aa}
\bibliography{Mecheri-d}
\end{document}